# Effective Spin-2 Quasi-particles at Linear Dispersive Five-fold Degenerate Points with Tunable Topological Chern Numbers

Cheng Feng[*], Qiang Wang, S. N. Zhu and Hui Liu[*1]

National Laboratory of Solid State Microstructures, School of Physics, Nanjing University, Nanjing 210093, China

**Abstract** In this work, which is based on spin-2 vectors and traceless spin-2 tensors, an effective Hamiltonian is constructed with a linearly dispersive five-fold degenerate point with spin-2 vector-momentum couplings. For the model without spin-2 vector-tensor coupling, the topological Chern numbers of five bands are calculated as 4, 2, 0, -2, -4. After including spin-2 vector-tensor coupling, separate topological Chern numbers are obtained for different couplings. A cubic lattice of atoms with five internal states is designed to realize two five-fold degenerate points. The Chern numbers of the bands can be changed by tuning the coupling coefficient. In this work we propose a theoretical design to obtain spin-2 quasi-particles.

## 1. Introduction

After Dirac discovered the spin of electrons, spin has played an ever increasing important role in modern physics. Physicists classify elementary particles by the spin property-statistics theorem of quantum statistics. Particles of half-integer spin that exhibit Fermi–Dirac statistics are called fermions, such as electrons, protons, and neutrons, with spin-1/2. Particles of integer spin, in other words full-integer, exhibit Bose–Einstein statistics and are called bosons. Examples of bosons are the Higgs boson with spin-0, the W, Z boson and photon with spin-1, and the graviton with spin-2 [1]. Although many particles predicted by physicists have been discovered experimentally, there are still many high-spin particles to be confirmed. Among them, the graviton with spin-2 is the particle predicted by quantum gravity theory, and the study of spin-2 particles will be of great value to the study of modern quantum gravity theory [2-6]. Experimentally confirming a single elementary particle requires very high energy. Some scholars tend to study the quasiparticles in condensed matter systems or artificial systems, which behave analogously to elementary particles. Remarkably, the discovery of relativistic Dirac fermions from graphene has had great impact in condensed matter physics [7]. Moreover, the Wely fermions [8-22] and the Dirac fermions [22-28], which are predicted in high-energy physics to have massless spin-1/2 vector, can emerge as quasiparticles in condensed systems or photonic crystals.

Furthermore, the triply degenerate points with spin-1 vectors have been theoretically and experimentally studied. In 2011 C. T. Chan et al. realized the triple degeneracy point with linear dispersion and zero-refractive-index materials through an accidental degeneracy in photonic crystals [29]. Then in 2016 they achieved Klein tunneling and super-collimation of pseudospin-1 electromagnetic waves in a two-dimensional photonic crystal [30]. In 2017 S. L. Zhu et al. obtained the triply degenerate point in an optical lattice using the pseudospin-1 Maxwell equation and studied the topological properties of Maxwell's quasiparticles [31]. Recently Zhang et al. theoretically developed new types of a triply degenerate point emerging from the coupling between a spin-1 vector and spin-1 tensor. These have distinct topological properties [32-34]. In addition to the works mentioned above, spin-1 can also be realized in condensed matter systems [35-39], quantum systems [40-42], and cold atom or lattice models [43-49]. As for spin-3/2, the massive relativistic spin-3/2 quasiparticle has been realized in condensed matter systems [50]. Moreover, it has been reported that superconductivity has been found in spin-3/2 topological semimetals [51-52] and optical lattices [53]. However, quasiparticles with high spin-2 are rarely studied. Inspired by this fact, here we study five-fold degenerate points with linear dispersion and a spin-2 vector.

---

[1] [*] Corresponding author.
*E-mail address:* liuhui@nju.edu.cn (Hui Liu), 1294924877@qq.com (Cheng Feng)

In this paper we first propose the simplest model to realize the five-fold degenerate point with linear dispersions. Then we investigate the topological properties emerging from the interplay of spin-2 vector and spin-tensor coupling. Finally, we realize two five-fold degenerate points in a cubic lattice model.

## 2. Five-Fold Degenerate Point

Normally, the Hamiltonian with spin-vector-momentum coupling $\sim \boldsymbol{k} \cdot \boldsymbol{S}$ can describe the linear dispersion with a degenerate point. For example, the linear dispersion of the Wely point can be depicted by $\sim \boldsymbol{k} \cdot \boldsymbol{\sigma}$ ($\boldsymbol{\sigma}$ is Pauli matrix) [8-14]. As for the triply degenerate point it can be described by the $\sim \boldsymbol{k} \cdot \boldsymbol{F}$ ($\boldsymbol{F}$ is the spin-1 vector) [30-34, 40, 49]. In this paper we propose a Hamiltonian with a spin-2 vector, which can produce a five-fold degenerate point with linear dispersion in all directions. The simplest five-fold degenerate point can be described by $H = \boldsymbol{k} \cdot \boldsymbol{S}$, with the spin-2 vector $\boldsymbol{S}$. Moreover, we can use commutation relations analogous to those of the orbital angular momentum [54] and spin creation and annihilation operators [55] to derive it. $S_x, S_y, S_z$ are depicted in equation (1).

$$S_x = \frac{1}{2}\begin{pmatrix} 0 & 2 & 0 & 0 & 0 \\ 2 & 0 & \sqrt{6} & 0 & 0 \\ 0 & \sqrt{6} & 0 & \sqrt{6} & 0 \\ 0 & 0 & \sqrt{6} & 0 & 2 \\ 0 & 0 & 0 & 2 & 0 \end{pmatrix}$$

$$S_y = \frac{1}{2i}\begin{pmatrix} 0 & 2 & 0 & 0 & 0 \\ -2 & 0 & \sqrt{6} & 0 & 0 \\ 0 & -\sqrt{6} & 0 & \sqrt{6} & 0 \\ 0 & 0 & -\sqrt{6} & 0 & 2 \\ 0 & 0 & 0 & -2 & 0 \end{pmatrix} \quad (1)$$

$$S_z = \begin{pmatrix} 2 & 0 & 0 & 0 & 0 \\ 0 & 1 & 0 & 0 & 0 \\ 0 & 0 & 0 & 0 & 0 \\ 0 & 0 & 0 & -1 & 0 \\ 0 & 0 & 0 & 0 & -2 \end{pmatrix}$$

The band dispersion is displayed in Fig. 1 (a), in which there is one flat band located at the center and four linear band dispersions along all directions intersecting at one point, i.e., five-fold degenerate points. The topological property of this degenerate point can be characterized by the first Chern number [56-57], which is proportional to the total variation of the Berry phase [58] as shown in equations (2) and (3).

$$\gamma_n = \oint \mathbf{\Omega}(\mathbf{k}) \cdot d\mathbf{S} \qquad (2)$$

$$C_n = \frac{\gamma_n}{2\pi}, \qquad (3)$$

where S is the closed surface surrounding the degenerate point, and the Berry curvature is $\mathbf{\Omega}(\mathbf{k}) = \nabla \times \mathbf{A}(\mathbf{k})$ with Berry connection $\mathbf{A}(\mathbf{k}) = -i\langle\psi_n(\mathbf{k})|\nabla_\mathbf{k}|\psi_n(\mathbf{k})\rangle$ defined by the wave function $\psi_n$ in the nth band (n=1, 2, 3, 4, 5). Here, we use the discretization algorithm to calculate the Berry phase of the five-fold degenerate point in spherical coordinates [59-60] as shown by equation (4):

$$\gamma_{n,\theta} = -\mathrm{Im}\sum_{i=1}^{N}\ln\langle\psi_{n,\theta,\varphi_i}|\psi_{n,\theta,\varphi_{i+1}}\rangle, \quad (4)$$

where $\varphi_i = \frac{2\pi}{N}i$, when $N$ has been chosen to be large enough such that $\gamma_{n,\theta}$ converges. The Berry Phase variation with $\theta$ is depicted in Fig. 1 (b).

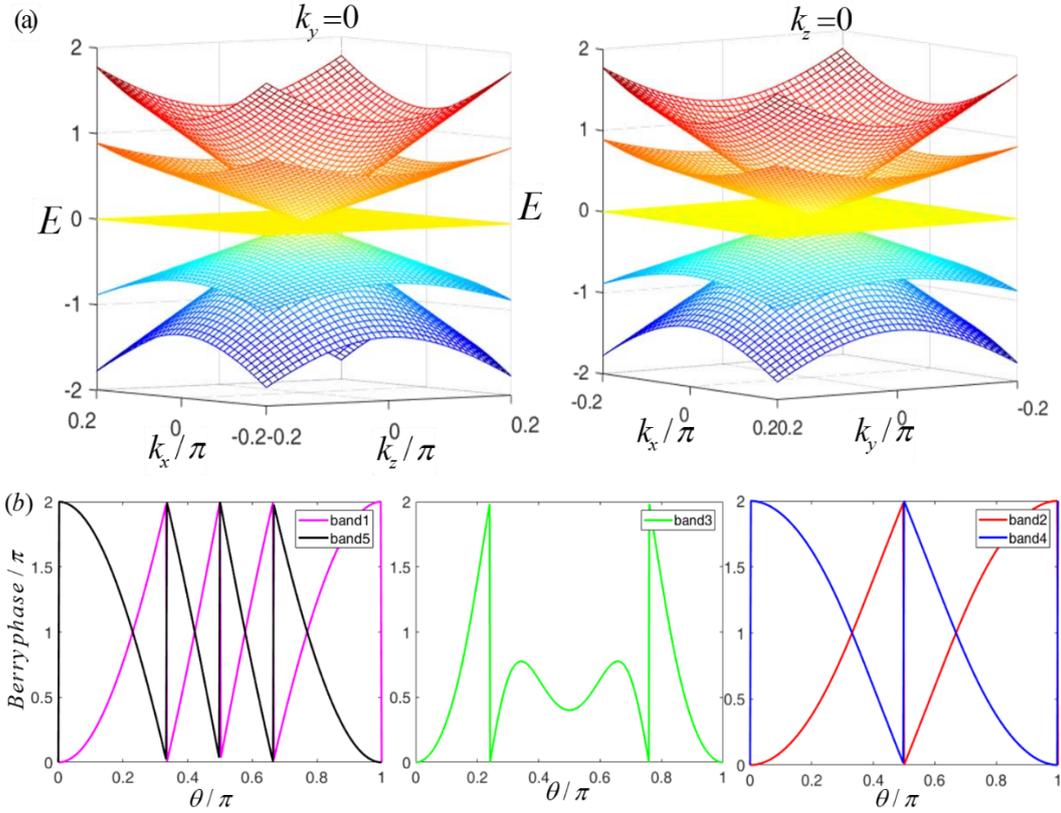

Fig. 1 (a) Band structure of the simplest model to form five-fold degenerate point $H = \mathbf{k} \cdot \mathbf{S}$. Fig. 1 (b) the Berry phase of the five-fold degenerate point calculated by wave functions of different bands.

The bands are labeled as band 1–5 from low to high. The total variation of band 1 and band 5 is $\pm 8\pi$, and for band 2 and band 4 it is $\pm 4\pi$, as for band 3 the change is zero. Therefore, the Chern numbers of the five-fold degenerate point formed by this simplest Hamiltonian are 4, 2, 0, -2, and -4.

### 3. New Topological Properties of the Five-Fold Degenerate Point

Inspired by the topological properties of the new types of triply degenerate points induced by the interplay of spin-1 vector-momentum couplings and spin-1 tensor-momentum couplings [32-34], we questioned whether there would be a five-fold degeneracy point with a new topology when the spin-2 tensor couplings are added to the simplest Hamiltonian. Thus, in addition to the three spin-2 vectors $S_x, S_y, S_z$, and in close analogy to the spin-1 tensor we derived a traceless spin-2 tensor $N_{ij} = S_i S_j + S_j S_i - \delta_{ij} S^2/5 (i,j = x,y,z)$. In this paper we report the topological properties of a new five-fold degenerate point formed by the interplay between spin-2 vector-momentum couplings and spin-2 vector-tensor couplings. The Hamiltonian of the model is given by equation (5) (the spin tensor is only coupled to one momentum $k_k$):

$$H(\mathbf{k}) = \mathbf{k} \cdot \mathbf{S} + \alpha k_k N_{ij} (i,j,k = x,y,z). \quad (5)$$

Here $\alpha$ is the coupling coefficient. By tuning the coupling coefficient and adopting different combinations of spin-2 tensor-momentum couplings, we find several topological Chern numbers of five-fold degenerate points different from those of the simple model. The results are shown in Table 1; the bands are labeled as Band 1–5 from low to high.

TABLE 1: Spin momentum coupling with one spin tensor; the number in the table is the topological Chern number numerically calculated by the wave functions of Band 1–5 from low to high energy.

|  | Band 1 | Band 2 | Band 3 | Band 4 | Band 5 |
|---|---|---|---|---|---|
|  | 4 | 2 | 0 | -2 | -4 |
| $k_x N_{xx}$, | 3 | 3 | 0 | -3 | -3 |
| $k_y N_{yy}$, | 3 | 2 | 0 | -2 | -3 |
| $k_z N_{zz}$, | 3 | 1 | 0 | -1 | -3 |
|  | 2 | -1 | 0 | 1 | -2 |
| $k_x N_{xz}, k_x N_{xy}$ | 4 | 2 | 0 | -2 | -4 |
| $k_y N_{xy}, k_y N_{yz}$ | 2 | 2 | 0 | -2 | 2 |
| $k_z N_{xz}, k_z N_{yz}$ | 0 | 0 | 0 | 0 | 0 |
| $k_x N_{yy}, k_x N_{zz}, k_x N_{yz}$<br>$k_y N_{xx}, k_y N_{zz}, k_y N_{xz}$<br>$k_z N_{xx}, k_z N_{yy}, k_z N_{xy}$ | 4 | 2 | 0 | -2 | -4 |

As illustrated in Table 1, no matter what value $\alpha$ has, the Chern numbers remain the same as in the simplest model (4, 2, 0, -2, -4) when the spin tensor $N_{ij}(i,j,k = x,y,z; i \neq k)$ is coupled to $k_k$. And when spin tensor $N_{ij}$ or $N_{ji}(i,j = x,y,z; i \neq j)$ to $k_i$; when $0 \leq |\alpha| \leq \frac{2}{3}$, the Chern numbers are 4, 2, 0, -2, -4; when $\frac{2}{3} < |\alpha| \leq 2$ the result is 2, 2, 0, -2, -2; when $|\alpha| > 2$, the result is 0. Moreover, when spin tensor $N_{ii}$ is coupled to $k_i(i = x,y,z)$, and $0 \leq |\alpha| \leq \frac{1}{3}$, the Chern numbers are 4, 2, 0, -2, -4; when $\frac{1}{3} < |\alpha| < \frac{1}{2}$, the result is 3, 3, 0, -3, -3; when $|\alpha| = \frac{1}{2}$, the Chern numbers are 3, 2, 0, -2, -3; when $\frac{1}{2} < |\alpha| \leq 1$, the Chern numbers are 3, 1, 0, -1, -3; when $|\alpha| > 1$, the result is 2, -1, 0, 1, -2.

We chose the spin tensor $N_{xz}$ coupling with $k_z$ as an example to illustrate the change in the topological Chern numbers. Here the Hamiltonian is defined as: $H(\mathbf{k}) = \mathbf{k} \cdot \mathbf{S} + \beta k_z N_{xz}$. We tune the coupling coefficient $\beta$ when calculating the band dispersion and the Chern numbers of this five-fold degenerate point. As displayed in Table 1, there are three different types of topological Chern numbers with varying $\beta$. So, we chose $\beta = 0.5, 1.2, 2.2$, and the band dispersions when $k_y = 0$ are depicted in Fig. 2 (a)-(c), respectively. Fig. 2 (d) illustrates that the topological Chern numbers change with the tuning of the coupling coefficient $\beta$. In particular when $|\beta| = \frac{2}{3}$ or $|\beta| = 2$, there will be degenerate lines in the direction $k_x = \pm k_z$.

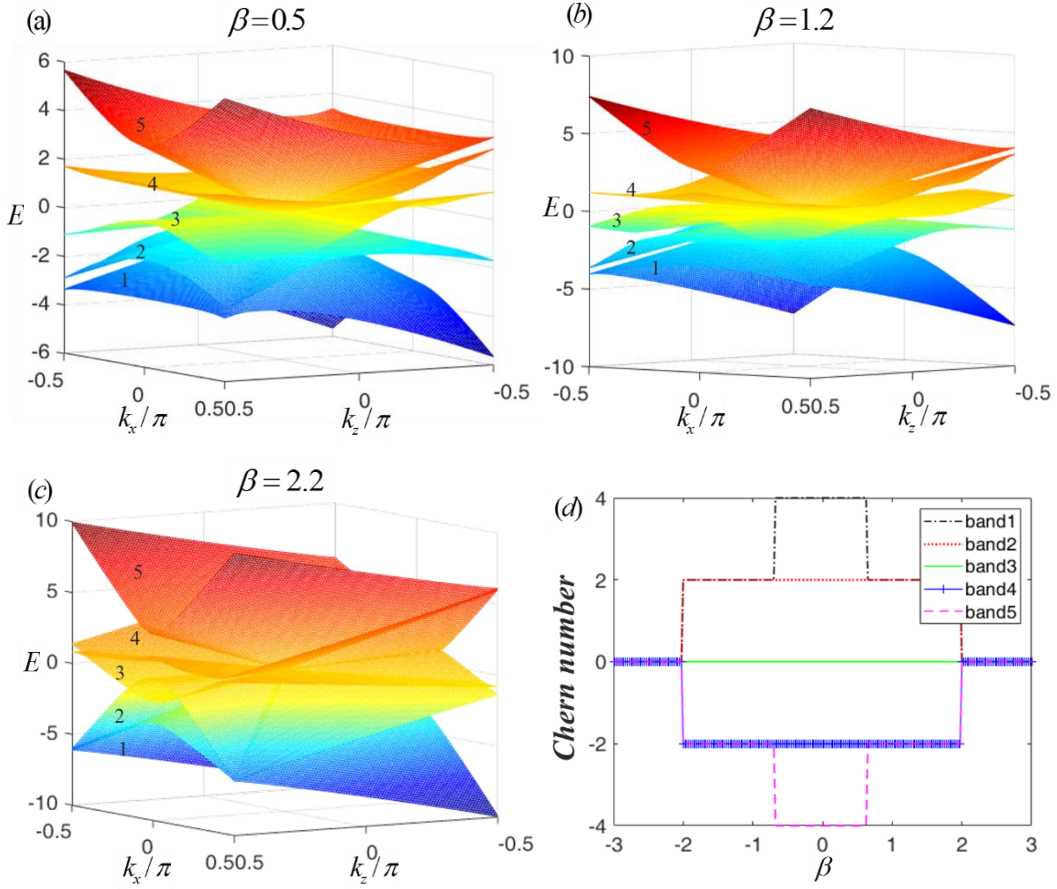

Fig. 2 (a)-(c) Band structure of $H(\mathbf{k}) = \mathbf{k} \cdot \mathbf{S} + \beta k_z N_{xz}$, as $\beta = 0.5, 1.2, 2.2$, and $k_y = 0$. Fig. 2 (d) is the phase diagram found by tuning $\beta$, from the highest band 5 (dashed line in magenta), second highest band 4 (+ blue line), middle band 3 (solid green line), the lowest band 1 (solid circle line in black), and the second lowest band 2 (dotted red line).

Further, we propose a tilted simple cubic lattice to obtain two five-fold degenerate points in $k$-space, and chose five atomic internal states to represent the five spin states $|2\rangle, |1\rangle, |0\rangle, |-1\rangle, |-2\rangle$. For the specific structure of this model, we considered non-interacting atoms in a titled cubic lattice, and chose five atomic internal states in the ground-states manifold to carry the spin states, as shown in Fig. 3(a). In addition, the atomic hopping configuration terms along the three coordinate axes are labeled by $T_{\pm x}, T_{\pm y}, T_{\pm z}$. Moreover, in our model the hopping term $T_{+x} \neq T_{-x}$ and $T_{+y} \neq T_{-y}$. The required broken symmetry could be achieved by titling the cubic lattice with the homogeneous energy gradient along the $x, y$ direction, which can be created by gravitation filed or the gradient of a dc- or ac-Stark shift [31].

The schematic diagram of hopping terms $T_{\pm x}, T_{\pm y}, T_{\pm z}$ along the three axes is illustrated in Fig. 3(b-f), in which we use five atomic internal states $|2\rangle, |1\rangle, |0\rangle, |-1\rangle, |-2\rangle$ to form the spin-2 basis. Along the three coordinate axes the hopping coefficient between spin states $|2\rangle$ to $\Delta_{\pm x,y,z}|2\rangle$ is $(2 + 2\gamma)t$, and between spin states $|1\rangle$ to $\Delta_{\pm x,y,z}|1\rangle$ is $(1 - \gamma)t$, and between spin state $|0\rangle$ to $\Delta_{\pm x,y,z}|0\rangle$ is $-2\gamma t$, and between spin state $|-1\rangle$ to $\Delta_{\pm x,y,z}|-1\rangle$ is $(-1 - \gamma)t$, and between spin state $|-2\rangle$ to $\Delta_{\pm x,y,z}|-2\rangle$ is $(-2 + 2\gamma)t$.

In addition, along the $x$ axes, the hopping coefficient between spin states $|1\rangle$ to $\Delta_{\pm x}|2\rangle$ and $|-2\rangle$ to $\Delta_{\pm x}|-1\rangle$ is $\mp it$, between the spin states $|0\rangle$ to $\Delta_{\pm x}|1\rangle$ and $|-1\rangle$ to $\Delta_{\pm x}|0\rangle$ is $\mp it\sqrt{6}/2$. At same time, along the $y$ axes, the hopping coefficient between spin states $|1\rangle$ to $\Delta_{\pm y}|2\rangle$ and $|-2\rangle$ to $\Delta_{\pm y}|-1\rangle$ is $\pm t$, between spin states $|0\rangle$ to $\Delta_{\pm y}|1\rangle$ and $|-1\rangle$ to $\Delta_{\pm y}|0\rangle$ is $\pm t\sqrt{6}/2$. Finally, we have to consider the on-site terms of each spin state, which are $-8(1+\gamma)t$, $-4(1-\gamma)t$, $8\gamma t$, $4(1+\gamma)t$, and $8(1-\gamma)t$ for spin states $|2\rangle, |1\rangle, |0\rangle, |-1\rangle, |-2\rangle$.

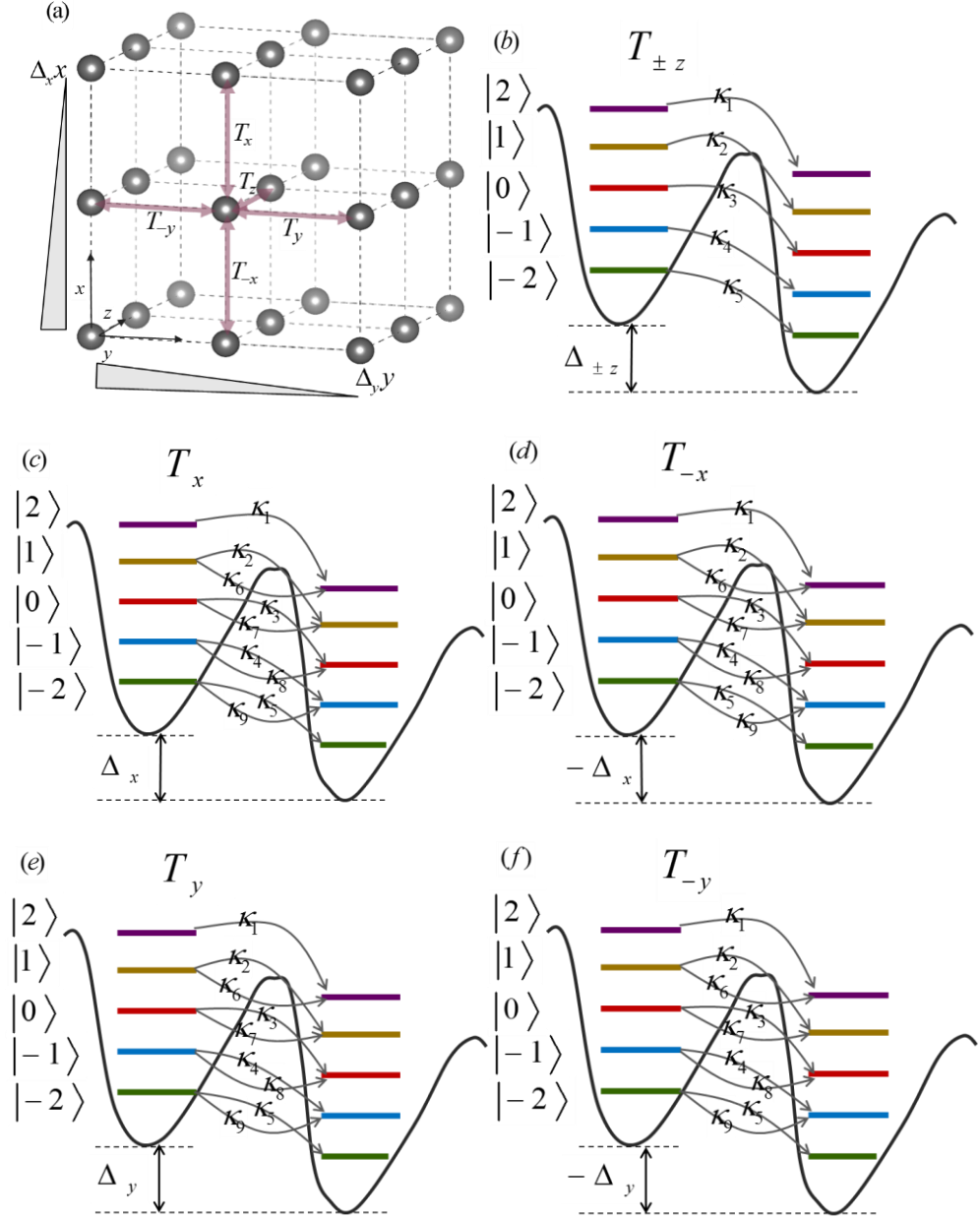

Fig. 3(a) the schematic diagram of the cubic lattice for the atomic hopping along the three axes denoted by $T_{\pm x}, T_{\pm y}, T_{\pm z}$. Fig. 3(b-f) the atomic hopping configuration for the model of equation (6) along the three axes, where the hopping coefficient $\kappa_i (i=1,2,3,4,5,6,7,8,9)$ is given in Table 2.

In order to illustrate the atomic hopping terms along three axes more concretely, the equation of $T_{\pm x}, T_{\pm y}, T_{\pm z}$ are given: $T_{\pm x,y} = (\kappa_1|2\rangle + \kappa_6|1\rangle) \to |2\rangle_{r\pm x,y} + (\kappa_2|1\rangle + \kappa_7|0\rangle) \to |1\rangle_{r\pm x,y} + (\kappa_3|0\rangle + \kappa_8|-1\rangle) \to |0\rangle_{r\pm x,y} + (\kappa_4|-1\rangle + \kappa_9|-2\rangle) \to |-1\rangle_{r\pm x,y} + \kappa_5|-2\rangle \to |-2\rangle_{r\pm x,y} + \text{H.c.}$,

$T_{\pm z} = \kappa_1 |2\rangle \to |2\rangle_{r\pm z} + \kappa_2 |1\rangle \to |1\rangle_{r\pm z} + \kappa_3 |0\rangle \to |0\rangle_{r\pm z} + \kappa_4 |-1\rangle \to |-1\rangle_{r\pm z} + \kappa_5 |-2\rangle \to |-2\rangle_{r\pm z} + H.c.$.

$T_{\pm z}$ denote that atom in superposition spin states $\kappa_1|2\rangle + \kappa_2|1\rangle + \kappa_3|0\rangle + \kappa_4|-1\rangle + \kappa_5|-2\rangle$ at site $r$ hopping to site $r\pm z$ change to superposition spin state $|2\rangle + |1\rangle + |0\rangle + |-1\rangle + |-2\rangle$. And $T_{\pm x,y}$ means that atom in superposition states $\kappa_1|2\rangle + (\kappa_2 + \kappa_6)|1\rangle + (\kappa_3 + \kappa_7)|0\rangle + (\kappa_4 + \kappa_4)|-1\rangle + (\kappa_5 + \kappa_9)|-2\rangle$ at site $r$ hopping to site $r\pm x,y$ change to another superposition spin states $|2\rangle + |1\rangle + |0\rangle + |-1\rangle + |-2\rangle$. Along different axes, the hopping coefficient $\kappa_1, \kappa_2, \kappa_3, \kappa_4, \kappa_5$ is remain same, but $\kappa_6, \kappa_7, \kappa_8, \kappa_9$ are variable which are display in Table 2.

TABLE 2: The hopping coefficient $\kappa_1, \kappa_2, \kappa_3, \kappa_4, \kappa_5, \kappa_6, \kappa_7, \kappa_8, \kappa_9$ among spin states $|2\rangle, |1\rangle, |0\rangle, |-1\rangle, |-2\rangle$ displayed in Fig.3 (b-f).

|  | $\kappa_1$ | $\kappa_2$ | $\kappa_3$ | $\kappa_4$ | $\kappa_5$ | $\kappa_6$ | $\kappa_7$ | $\kappa_8$ | $\kappa_9$ |
|---|---|---|---|---|---|---|---|---|---|
| $T_x$ | $(2+2\gamma)t$ | $(1-\gamma)t$ | $-2\gamma t$ | $(-1-\gamma)t$ | $(-2+2\gamma)t$ | $-it$ | $-it\sqrt{6}/2$ | $-it\sqrt{6}/2$ | $-it$ |
| $T_{-x}$ | $(2+2\gamma)t$ | $(1-\gamma)t$ | $-2\gamma t$ | $(-1-\gamma)t$ | $(-2+2\gamma)t$ | $it$ | $it\sqrt{6}/2$ | $it\sqrt{6}/2$ | $it$ |
| $T_y$ | $(2+2\gamma)t$ | $(1-\gamma)t$ | $-2\gamma t$ | $(-1-\gamma)t$ | $(-2+2\gamma)t$ | $t$ | $t\sqrt{6}/2$ | $t\sqrt{6}/2$ | $t$ |
| $T_{-y}$ | $(2+2\gamma)t$ | $(1-\gamma)t$ | $-2\gamma t$ | $(-1-\gamma)t$ | $(-2+2\gamma)t$ | $-t$ | $-t\sqrt{6}/2$ | $-t\sqrt{6}/2$ | $-t$ |
| $T_{\pm z}$ | $(2+2\gamma)t$ | $(1-\gamma)t$ | $-2\gamma t$ | $(-1-\gamma)t$ | $(-2+2\gamma)t$ | -- | -- | -- | -- |

Then the Hamiltonian of this well-designed lattice model can be written as equation (6):

$$H(k) = 2t \sin k_x S_x + 2t \sin k_y S_y + 2t(S_z + \gamma N_{zz})(\cos k_x + \cos k_y + \cos k_z - 2). \quad (6)$$

In this equation (6), the parameter $\gamma$ is the exactly same as the parameter $\gamma$ of atomic hopping coefficient of the cubic lattice between spin states $|i\rangle$ to $\Delta_{\pm r}|i\rangle$ ($i$=-2,-1,0,1,2), which is exist in $\kappa_1, \kappa_2, \kappa_3, \kappa_4, \kappa_5$ in Fig. 3(b-f). As the tuning of parameter $\gamma$ in the equation (6), the hopping coefficient between spin states along three axes will change equally.

In this model, the two five-fold degenerate points exist at $(0,0,\pm\frac{\pi}{2})$. Around these two points, the Hamiltonian can be expanded as $H(\pm\delta k) = 2t(\delta k_x S_x + \delta k_y S_y \mp \delta k_z S_z \mp \gamma \delta k_z N_{zz})$, this effective Hamiltonian is one specific type in Table 1.

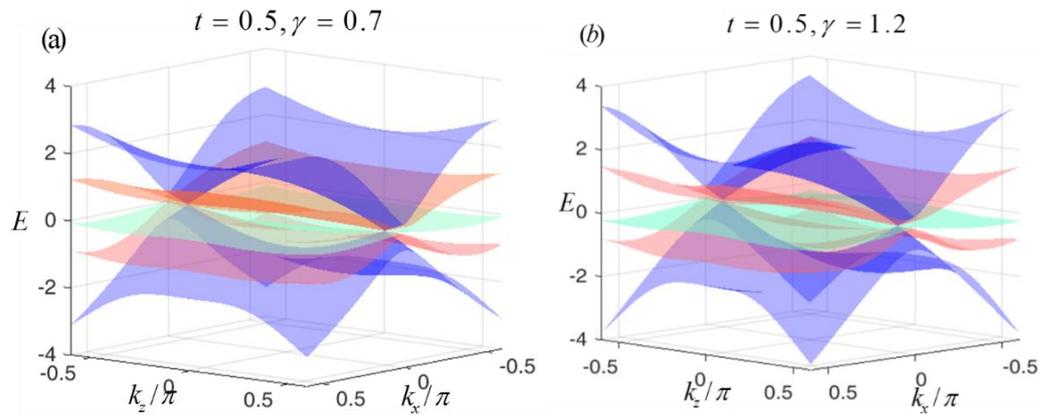

Fig. 4 (a-b) Band structure with two five-fold degenerate points when $t = 0.5, \gamma = 0.7$ and $t = 0.5, \gamma = 1.2$ respectively.

As displayed in Table 1, because the spin tensor $N_{zz}$ is coupled to $k_z$, there is a change in Chern numbers as the coupling coefficient $\gamma$ is tuned. When $\frac{1}{2} < |\gamma| \leq 1$, the Chern numbers are

3, 1, 0, -1, -3; when $|\gamma| > 1$, the result is 2, -1, 0, 1, -2. The band dispersion for different types of Chern numbers is illustrated in Fig. 4 (a)-(b). The parameter in Fig. 4 (a) is $t = 0.5, \gamma = 0.7$; in Fig. 4 (b) it is $t = 0.5, \gamma = 1.2$. From these two figures we can clearly see that there are two five-fold degenerate points at $(0,0,\pm\frac{\pi}{2})$. Then we calculate the Chern numbers at $(0,0,-\frac{\pi}{2})$, by tuning the coupling coefficient $\gamma$; when $t = 0.5, \gamma = 0.7$, the Chern numbers are -3, -1, 0, 1, 3; when $t = 0.5, \gamma = 1.2$, the Chern numbers are -2, 1, 0, -1, 2. At the same time, the other degenerate point at $(0,0,\frac{\pi}{2})$, when $t = 0.5, \gamma = 0.7$, gives the Chern numbers as 3, 1, 0, -1, -3; and when $t = 0.5, \gamma = 1.2$, the Chern numbers are 2, -1, 0, 1, -2. Thus, these two points have exactly opposite Chern numbers when tuning $\gamma$. In this model we can achieve five types of Chern numbers of two five-fold degenerate points displayed in Table 1 only by tuning the coupling coefficient $\gamma$.

Theoretically, we use the five atomic internal states to encode the spin states $|2\rangle, |1\rangle, |0\rangle, |-1\rangle, |-2\rangle$ to obtain two five-fold degenerate points in the cubic lattice model, and by tuning the parameter $\gamma$ to adjust the hopping coefficient between spin states, we can achieve different sets of topological Chern numbers.

## 4. Conclusion

This work proposes a theoretical model to achieve linearly dispersive five-fold degenerate points with different topological Chern numbers through the interplay between spin vector-momentum couplings and spin tensor-momentum couplings. A tilted simple cubic lattice is proposed to realize two linearly dispersive five-fold degenerate points in *k* space, in which we use the atomic internal ground states to represent the spin-2 states $|2\rangle, |1\rangle, |0\rangle, |-1\rangle, |-2\rangle$. The topological properties of the degenerate point can be tuned just by the coupling coefficient. Of course, our research also has some limitations. For example, we can consider two or three spin vector-tensor coupling cases. In this case we may get new five-fold degenerate points with special topological properties. At the same time, our paper just provide a theoretical proposal of the cubic lattice for the implementation of the Hamiltonian. We ignore the difficulty of realization of spin-dependent tunneling in the real experiment. By now, spin-momentum coupling just propose and realize for spin-1 and spin-1/2 system in the ultracold atomic gas and optical lattice [61-75]. A scheme for realizing a spin-tensor-momentum coupling of spin-1 atoms has also been proposed recently [32].Spin-vector-momentum coupling for spin-2 are rarely studied in condensed-matter physics or artificial systems. We believe that with the development of experimental platform and experimental techniques, fine-tuning of the spin-dependent tunneling will be possibly to implement experimentally. Furthermore, we will advance our theoretical model to make it more feasible and operational for the experiment in the future.

**Acknowledgements**

This work was financially supported by the National Key R&D Program of China (2017YFA0303702 and 2017YFA0205700) and the National Natural Science Foundation of China (11690033, 11621091, and 61425018).